# Johnson-Kendall-Roberts theory applied to living cells

**Yeh-Shiu Chu[#], Sylvie Dufour[#], Jean Paul Thiery[#], Eric Perez[✛] and Frédéric Pincet[✛]**

[#]UMR 144, Centre National de la Recherche Scientifique et Institut Curie, 26 rue d'Ulm, 75248 Paris Cedex 05, France. [✛]Laboratoire de Physique Statistique de l'Ecole Normale Supérieure, UMR8550, Centre National de la Recherche Scientifique et Universités Paris 6 et 7, 24 rue Lhomond, 75251 Paris Cedex 05, France.

Address all correspondence to : Frederic Pincet: Laboratoire de Physique Statistique de l'Ecole Normale Supérieure, UMR8550, 24 rue Lhomond, 75251 Paris Cedex 05, France. Tel: (33) 1 44 32 25 02, Fax: (33) 1 44 32 34 33, e-mail: Frederic.Pincet@lps.ens.fr

**Abstract**

JKR theory is an accurate model for strong adhesion energies of soft slightly deformable material. Little is known about the validity of this theory on complex systems such as living cells. We have addressed this problem using a depletion controlled cell adhesion and measured the force necessary to separate the cells with a micropipette technique. We show that the cytoskeleton can provide the cells with a 3-D structure that is sufficiently elastic and has sufficiently low deformability for JKR theory to be valid. When the cytoskeleton is disrupted, JKR theory is no longer applicable.

A quantitative understanding of the adhesion of living cells is not often possible and the study reported here is one of the rare exceptions. In contrast, the adhesion of solid elastic bodies has been extensively studied in the past and a complete mathematical description has been derived[1]. In general, when the contacting surfaces adhere only weakly and deform little, the DMT approach[2] allows prediction of the behavior of the system. At higher adhesion and deformability, when adhering surfaces are subject to a separating force, there is a finite, non-zero contact area at separation. In this case, JKR theory[3] gives the relation between the pull off force $F_s$ and the adhesion energy $W_{adh}$ via the radii of curvature of the materials. For solid, homogeneous spheres, this relation becomes:

$$W_{adh} = 2F_s/(3\pi R_m) \qquad\qquad (1)$$

Where $R_m$ is the geometrical mean of the radii of the two spheres.

Many experimental studies on simple elastic materials have verified this description[4]. Similarly, the relation between $F_s$ and $W_{adh}$ has been derived for spherical shells by Brochard and de Gennes[5]:

$$W_{adh} = F_s/(\pi R_m) \qquad\qquad (2)$$

However, the adhesion of soft bodies such as cells is much more difficult to characterize. Several attempts to probe the adhesion strength of two biological cells have been made using techniques including shear flow or centrifugation[6]. Adhesion experiments using micromanipulation were conducted more than a decade ago[7,8] using red blood cells, which have well-defined membrane elasticity and a relatively simple, liquid interior. In contrast, it is much more difficult to extract quantitative results from adhesion measurements involving nucleated cells, which are often characterized by an irregular surface with folds and wrinkles and whose interior exhibits a complex rheology. Chien's group has developed a model inspired by Evans' results[9] involving the mechanical equilibrium of the cell membrane. Using this model, they measured the adhesion between cytolytic T cells and target cells[10,11]. Treating the separation of the cells as a peeling process, they analyzed their experiments in terms of adhesion energies and junction avidity.

The present study involves living cells which do not spontaneously adhere. We cause them to adhere through a depletion effect in the suspending medium. We show that, when the cytoskeleton of the cells has a complete 3-D structure that maintains a slightly deformable spherical shape, JKR theory is applicable to relate the separation force to the adhesion energy. It gives an elastic modulus coherent with the one independently measured with a surface force apparatus and with those found in the literature[12]. In this case where the 3-D cytoskeleton is responsible for their spherical shape, the cells do not behave like shells but like solid elastic spheres.

The general principle of our approach consists of micromanipulating two murine sarcoma S180 cells[13] with micropipettes, making them adhere in a highly concentrated dextran solution and balancing the depletion-induced adhesion by the aspiration pressure in a micropipette.

It is well documented that non-adsorbing, water-soluble polymers can induce an attraction of phospholipid bilayers[14-15]. The adhesion energy $W_{adh}$ induced by the depletion of dextran has been measured experimentally on lipid vesicles[16] and analyzed theoretically[17]. De Gennes has derived the expression of $W_{adh}$ as a function of the volume fraction of polymers $\phi$:

$$W_{adh}=(k_BT/a^2)\phi^{1.5} \qquad\qquad\qquad (3)$$

where $k_BT$ is the thermal energy and a the size of a monomer.

For this study, we used a protocol similar to that used by Chien's group[11]. It is described in figure 1. Before analyzing the adhesion behavior, we establish that the adhesion observed in polymer solution is due only to this depletion effect. It was already known that S180 cells are devoid of intrinsic intercellular adhesion properties[18] because they do not express cell-cell adhesion receptors at their surface. This is consistent with our observation that S180 cells brought to close contact do not adhere without dextran. In contrast, in the presence of dextran, S180 cells do adhere when they are mechanically pushed together with the micropipettes. Equilibrium under zero compression is reached after this mechanical constraint is removed (after less than a second). Further, the observation that adhering cells separated immediately after transfer in a dextran-free chamber shows that no receptor was activated during the adhesion phase. This indicates that the adhesion of S180 cells observed here was purely a depletion effect.

During separation, the cells appear elastic and slightly deformable (see figure 1) and the contact area at separation remains finite. Therefore, it is interesting to analyze the separation process with JKR theory and with the spherical shell model. For this purpose, it is necessary to measure the pull-off force.

As shown by Yeung and Evans[19], the cells may display viscoelastic behaviors that could induce force gradients. To avoid any artifact due to this problem, we have checked that the aspiration force in the pipette equals the force transmitted to the contact zone: we used a direct method of probing this transmitted force by aspirating a cell in a 4-5 μm micropipette with a gentle suction and placing the opposite side of the cell on a spring (a microneedle with a known stiffness), the results of these force experiments indicate that, in the range of force, time and velocity used, the measured force equals exactly the aspiration one.

Thus, it is possible to test the validity of JKR and spherical shell theories on these cells. The separation force $F_s$ is close to the average of the aspiration forces of the penultimate cycle n-1 and the last cycle n:

$$F_s = \pi(\Delta P_{n-1} + \Delta P_n)\, R_p^2 \,/2 \qquad\qquad (4)$$

where $R_p$ is the pipette inner radius, $\Delta P_i$ being the aspiration during cycle i.

The adhesion energy predicted by JKR and spherical shells theories can be calculated from the measurements of $F_s$ and the radii of the cells. These values can be compared (figure 2) to the theoretical[17] expression for the energy due to the depletion effects (eqn 3) and the experimental measurements[16] of that energy. Figure 3 shows a very good agreement with JKR theory while spherical shells theory does not seem to be suitable.

To check that JKR theory is indeed valid, we have measured the variation of the contact area $R_c$ during the separation process and deduced the elastic modulus K from the relation[3]:

$$(R_c)^3 = \frac{R_m}{2K}\left[-F + 3\pi R_m W_{adh} + \sqrt{-6\pi R_m W_{adh}F + \left(3\pi R_m W_{adh}\right)^2}\right] \qquad\qquad (5)$$

where F is the (positive) pulling force. The results are plotted in figure 3 and give: K= 3500 +/- 1500 Pa. To check this value, we have conducted Surface Force Apparatus[20] experiments between two layers of cells in which the reduction of the two layers thickness with the load is measured (figure 4). These measurements give: K= 4200 +/- 1000 Pa which is in excellent agreement with values obtained by micromanipulation and with values from the literature[12] (1-5 kPa). As a final proof of the validity of JKR theory, the ratio between the contact radius at separation and the one under zero load was measured. The obtained value is 0.65 +/- 0.12, again in excellent agreement with the expected one, 0.63. Therefore, the main features of JKR theory are verified here. This result may seem surprising as living cells in general display very complex mechanical behaviors and JKR should obviously not be valid for all types of cells. In the present case, the cytoskeleton is responsible for the shape of the cells and its 3-D elastic properties. We have verified by imaging actin, tubulin and vimentin filaments that the S180 cells have an extended 3D-cytoskeleton (data not shown). However, elasticity is only expected of the behavior of the cytoskeleton for shape changes that are sufficiently rapid that there is no time for the cytoskeleton to exhibit plastic flow during the detachment. This is the case here. These experiments lasted a few tens of seconds, whereas the time taken by a cell to regain its spherical shape after it has been expelled from a pipette was a few minutes.

To confirm the assumption that the cytoskeleton is responsible for the elastic behavior of the cell, the same micromanipulation experiments were done in the presence of 0.1 µM or 1.5 µM of Latrunculin B (Lat) which inhibits actin polymerization and sequesters actin monomers[21,22]. When the cell is made more deformable by alteration (0.1 µM Lat) or disruption (1.5 µM Lat) of the actin cytoskeleton network, there is a drastic change in the adhesion measurements as shown in Figure 5. In the first concentration, JKR theory seems to work correctly at low dextran concentrations (weak forces) while it is not applicable at higher ones. In 1.5 µM Lat, the measured apparent adhesion is weak and independent of the dextran concentration. In these cases, the cells present a much larger deformation and take a long time (up to several minutes) to recover their initial shape and it is meaningless to try to deduce adhesion energies with the approach presented here. The actin cytoskeleton is mostly cortical in round cells in suspension and allows the mechanical connection of the membrane to the tridimensional elastic structure of the rest of the cell. It is therefore not surprising that in this case JKR and spherical shell theories are not valid anymore.

These measurements show that JKR theory can reasonably be applied to predict the adhesion energy of these cells. Micropipette experiments are ideal to measure such an adhesion as the aspiration pressure gives a good measurement of the separation force. Whether such measurements are valid for cells of other kinds remains open. The applicability of JKR theory to the adhesion of other living cells could be checked directly using depletion forces, as here. However these results suggest that the deformation of the cell during the detachment process is a good indicator of whether JKR or spherical shells theories are applicable: if the cells present a small deformation with a finite contact area at separation, this suggests a nearly elastic behavior of the cytoplasm and therefore the likelihood that these theories will be applicable.

**Acknowledgments**

The authors wish to thank Professor de Gennes for sharing his insights on depletion-induced adhesions.

This work was supported by the Centre National de la Recherche Scientifique, The Institut Curie (Programme Incitatif et Coopératif, Physicochimie des structures biologiques complexes), the Association pour la Recherche sur le Cancer (grant N° 5653) and the Fifth Framework Program from the European Community (grant N° QLG11-CT-2001-00869). Y.-S. Chu benefited from a France-Taiwan Ministry of Foreign Affairs PhD fellowship.

References

1. D. Maugis, J. Colloid Interf. Sci. **150**, 243 (1992). B.D. Hughes, L.R. White, J. Chem. Soc. Farad. Trans. 1 **76**, 963 (1980).

2. B.V., Derjaguin, V.M. Muller, and Y.P. Toporov, J. Colloid Interf. Sci. **53**, 314 (1975).

3. K.L. Johnson, K. Kendall, and A.D. Roberts. Proc. R. Soc. Lon. Ser. A. **324**, 301 (1971).

4. J.N. Israelachvili, E. Perez, and R.K. Tandon, J. Colloid Interface Sci. **78**, 260 (1980). E.D. Shchukin, in *Microscopic aspects of adhesion and lubrification* edited by J.M. Georges, 389-402, (Elsevier, Amsterdam, 1982). E.D. Shchukin, E.A. Amelina, and V.V. Yaminsky, Colloids Surf. **2**, 221 (1981)

5. F. Brochard-Wyart, and P.G. de Gennes, C.R. Physique, **4**, 281 (2003)

6. A.S.G. Curtis, J.M. Lackie, *Measuring Cell adhesion* (John Wiley & Sons, New York, 1990).

7. N. Mohandas, and E. Evans, J. Clin. Invest **76**,1605 (1985).

8. E. Evans, D. Berk, A. Leung, and N. Mohandas, Biophys. J., **59**, 849 (1991).

9. E. Evans, and A. Leung, J. Cell Biol. **98**,1201 (1984).

10. K.-l. P. Sung, L. A. Sung, M. Crimmins, S. J. Burakoff, and S. Chien, Science **234**,1405 (1986).

11. A. Tozeren, K.-l. P. Sung, and S. Chien, Biophys. J. **55**, 479 (1989).

12. M. S. Turner, and P. Sens. *Biophysical Journal* **76**, 564 (1999).

13. For the experiments, the same type of cells was used: murine sarcoma S180 cell. S180 cell cultures were trypsinized with 0.05% trypsin, 0.02% EDTA (In Vitrogen) and seeded at sufficiently high density to yield a nearly confluent culture 18 to 24 h later. At that point, cells were treated with 0.01% trypsin in HMF (magnesium-free HEPES-based buffer containing 10 mM calcium, pH 7.4). After centrifugation, the cell suspension was pipetted gently in HCMF (calcium- and magnesium-free, HEPES-based) to yield isolated cells.

Micromanipulation measurements were performed on a Leica inverted microscope at 37°C. All surfaces in contact with the cells were precoated with bovine serum albumin (10 mg/ml; fraction V, Sigma). Before the assay, the chamber was charged with $CO_2$-independent medium (In Vitrogen) supplemented with 1% FCS, penicillin (100 IU/ml) and streptomycin (100 µg/ml). Pipettes were filled with sterile isotonic sucrose solution (300-330 mOs) and preincubated in bovine serum albumin. The pressure sensor to control the aspiration pressure was a Validyne, model DP103-38, ranging from 0 to 50000 Pa.

14. E. Evans, D. J. Klinggenberg, W. Rawicz, and F. Szoka, Langmuir **12**:3031-3037 (1996).

15. E. Evans, and B. Kukan, Biophys. J. **44**, 255 (1983).

16. E. Evans, and D. Needham, Macromolecules **21**, 1822 (1988).

17. P-G. de Gennes, in *Scaling Concept in polymer physics*, Cornell University Press, Ithaca and London, second edition (1985).

18. D. R. Friedlander, R.-M. Mege, B. A. Cunningham, and G. M. Edelman, Proc. Natl. Acad. Sci. USA. **86**, 7043 (1989).

19. A. Yeung, and E. Evans. *Biophys. J.* **56**, 139 (1989).

20. J. Israelachvili, and G.E. Adams. *J. Chem. Soc. Faraday Trans. 1* **74**, 975 (1978).

21. I. Spector, N. R. Shocket, D. Blasberger, and Y. Kashman. *Cell Motil. Cytoskeleton.* **13**, 127 (1989).

22. G. Segal, W. Lee, P.D. Arora, M. McKee, G. Downey, and C.A. McCulloch. *J. Cell Sci.* **114**, 119 (2001).

**Figure legends**

*Figure 1:*

(a and b) Two cells, held under weak aspiration by micropipettes, are placed in contact and one second later became adherent.

Separation process (c, d, e): One cell is held by micropipette B under strong aspiration. The aspiration applied to the other cell is increased and the micropipette A displaced away. Either the cell leaves micropipette A (c) or both cells separate (e). In the first case, the cell is resized by the micropipette A (d), the aspiration incremented and the micropipette A displaced again. This cycle is repeated until the cells separate and the separation force is deduced from the last aspiration pressures. During the measurements, the pipettes were moved at a velocity of about $20\mu m/s$. The whole process of separation lasted one minute at most. The aspiration level on pressure employed in each cycle was monitored continuously.

*Figure 2:*

Adhesion energy, as deduced from JKR theory (diamonds, eqn 1) and spherical shells (cross, eqn 2) as a function of the volume fraction of dextran. Two sizes of dextran molecules ($4.6 \cdot 10^5$ and $2 \cdot 10^6$ M.W) were used . The results can be compared to the theoretical ones given by de Gennes[16] (line) and to experimental ones obtained by Evans by contact angle measurements on lipid vesicles[15] (squares).

*Figure 3:*

Parameter $A = \dfrac{R_m}{2}\left[ -F + 3\pi R_m W_{adh} + \sqrt{-6\pi R_m W_{adh} F + \left(3\pi R_m W_{adh}\right)^2} \right]$ plotted as a function of $R_c^3$. As indicated in eqn 5, in the case of JKR theory, the slope gives the elastic modulus. The large error bars are due to the low accuracy in the measurement of the contact radius in optical microscopy. The points are taken from three different experiments at various dextran volume fractions.

*Figure 4:*

Force between two layers of S180 cells deposited on mica surfaces in a Surface Force Apparatus as a function of the parameter $(\delta^3 R)^{1/2}$ where $\delta$ is the reduction of the two cell layers thickness under compression and R the radius of the substrate. For δ smaller than the cell size, the slope gives the elastic modulus[3].

*Figure 5:*

(a, b and c) Morphology of the cells treated with Lat B during the separation process. Note the difference with figure 1. (d) Adhesion energy as it would be obtained through JKR theory (eqn 1) as a function of the dextran volume fraction in the presence of 0.1 μM (filled diamonds) or 1.5 μM (empty diamonds) latrunculine B. The solid line is the expected value deduced from the applied depletion force[16].

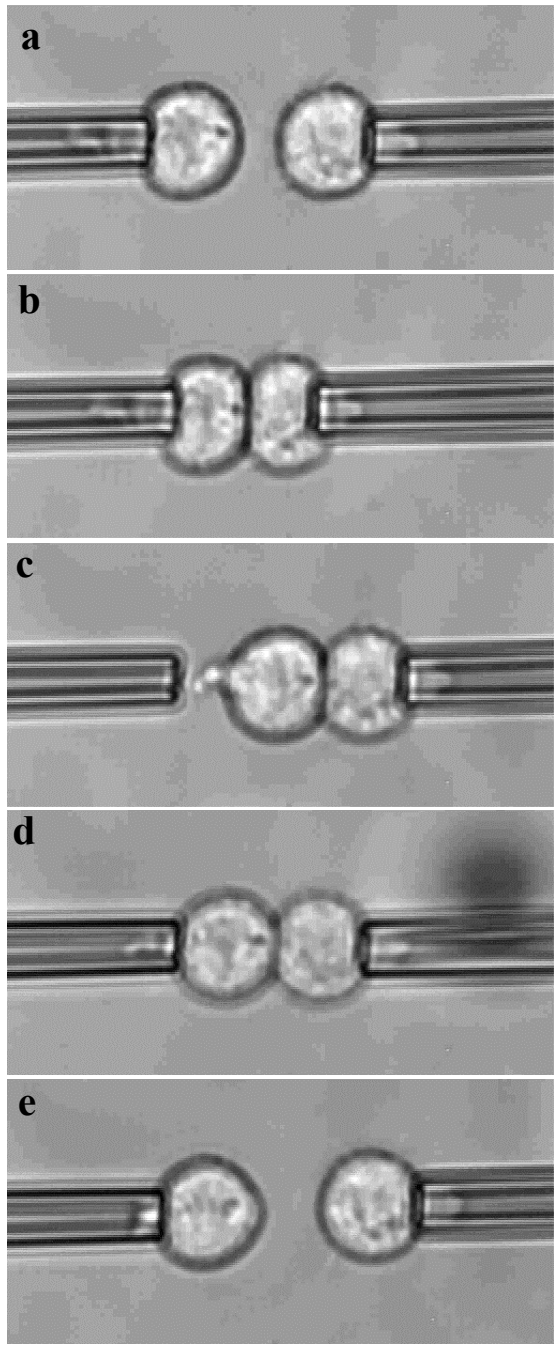

Figure 1

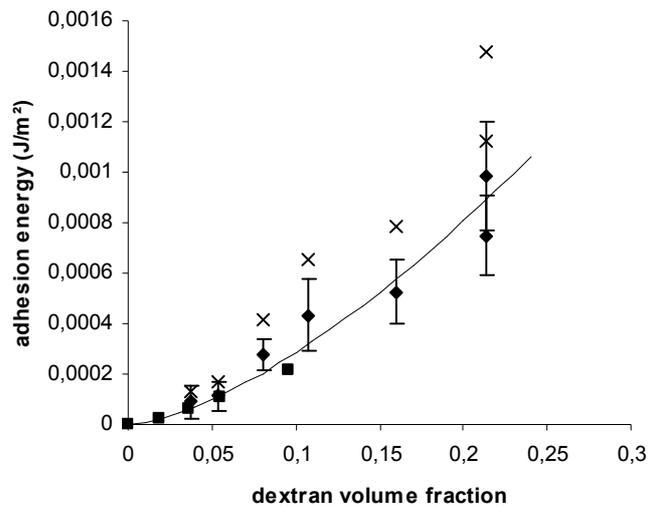

Figure 2

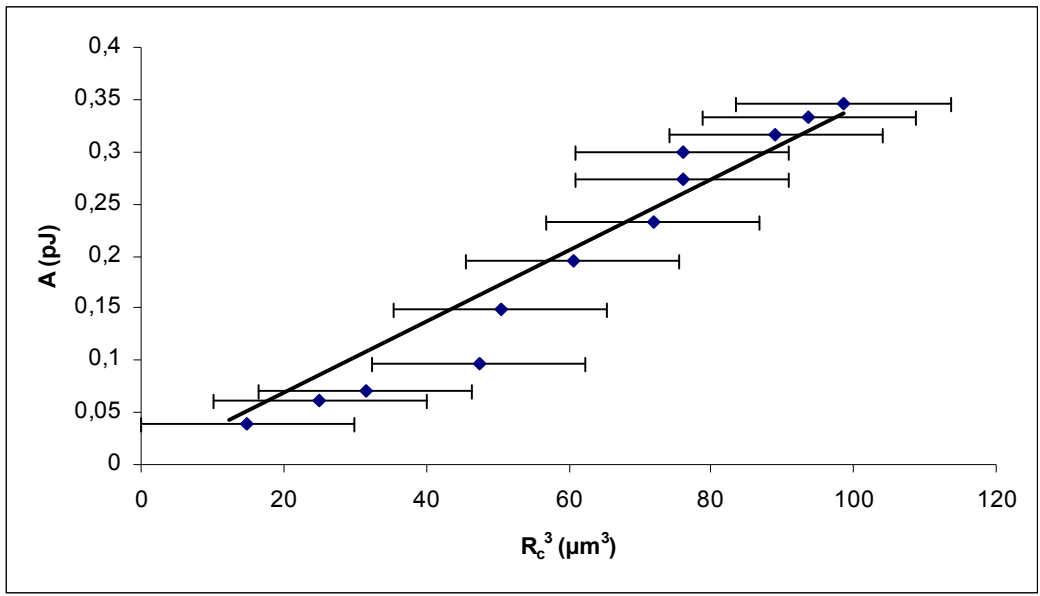

Figure 3

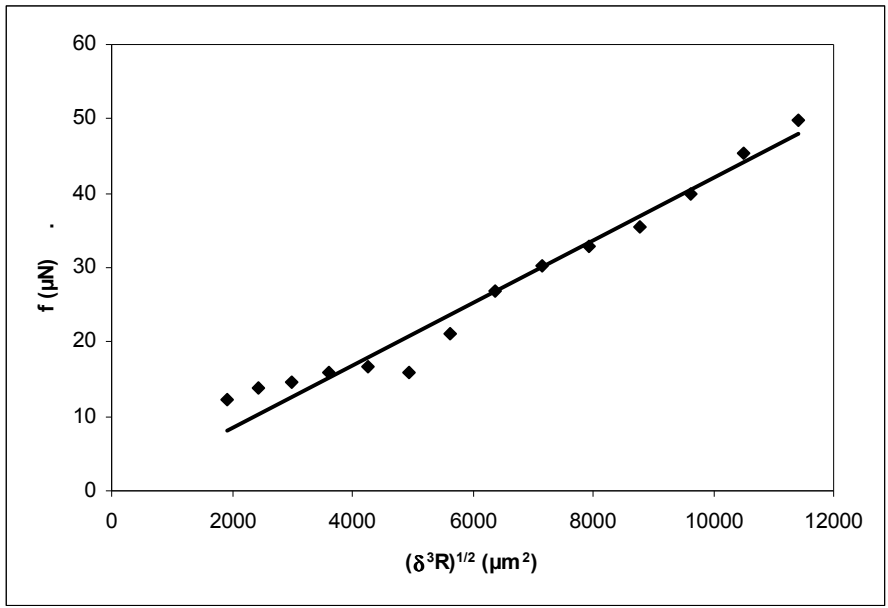

Figure 4

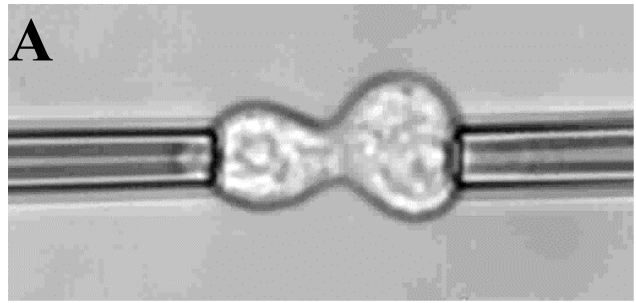

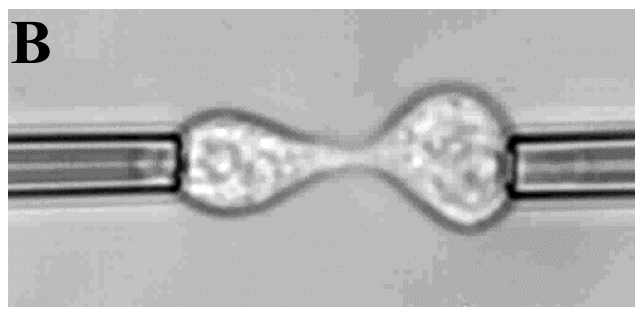

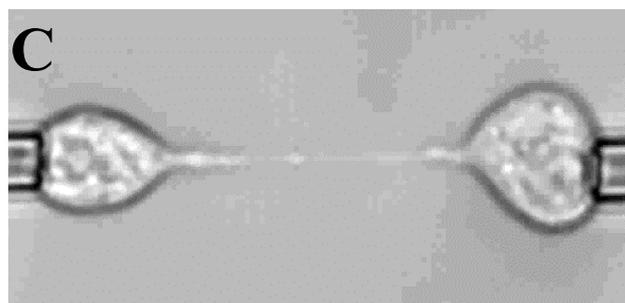

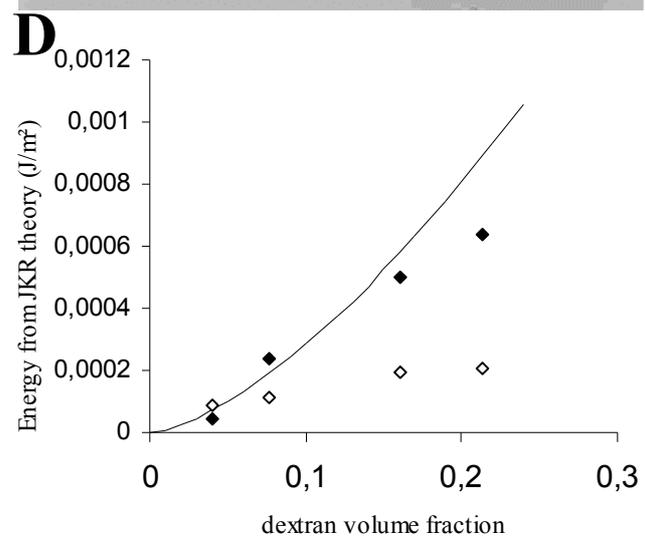

Figure 5